\documentclass[a4paper,12pt,twoside]{article}
\usepackage{graphicx,axion_string}

\begin{document}
\makeatletter
\@addtoreset{equation}{section}
\makeatother
\def\sltwo{{\sl sl}(2)}
\title{\sc\Large Non-Abelian Statistics Of Axion Strings}
\author{\large 
Masatoshi Sato$^{\dagger 1}$\\[7pt]
\sc $^\dagger$The Institute for Solid State Physics\\[3pt]
\sc The University of Tokyo, Kashiwanoha 5-1-5, \\[3pt]
\sc Kashiwa-shi, Chiba 277-8581, Japan}
\footnotetext[1]{msato@issp.u-tokyo.ac.jp}
\maketitle

PACS: 11.15Kc, 11.27+d

keywords: non-abelian statistics, axion, fermionic zero mode, 
Majorana-Weyl fermion. 

\vspace{-12cm}
\rightline{}
\vspace{11.5cm}
\thispagestyle{empty}

\begin{abstract}
We examine an axion string coupled to a Majorana fermion.
It is found that there exist a Majorana-Weyl zero mode on the string.
Due to the zero mode, 
the axion strings obey non-abelian statistics.  
\end{abstract}

Fermionic zero modes on topological defects often 
cause non-trivial phenomena in relativistic quantum field
theory. 
Some examples are fermion number fractionalization \cite{JR},
baryon number violation in the standard model \cite{tHooft},
superconductivity on strings \cite{Witten}, monopole catalysis of
proton decay \cite{Rubakov,Callan}, and so on.  
Geometrical relations between shapes of strings and fermion
number violation due to zero modes on the strings were 
investigated in Ref.\cite{Sato}.  
In this paper, we will present a novel phenomenon in relativistic
quantum filed theory which comes from fermionic zero
modes. 
{\it Fermionic zero modes cause non-abelian statistics of
strings in two spatial dimensions.}

In two spatial dimensions, exotic
generalizations of Fermi and Bose statistics are possible.
Wave functions in two spatial dimensions are 
representations of the braid group under interchanges of identical
particles.   
If the wave functions are vectors and the representation is given by
non-abelian matrices, the particle is said to obey non-abelian
statistics.
A particular type of non-abelian statistics is realized by the non-abelian
vortices that occur in theories with spontaneously broken non-abelian
symmetries 
\cite{WW,Bucher,BvDdW,LP}.  
Axion strings we examine here are excitations in
a theory with a spontaneously broken U(1) symmetry.   

In condensed matter physics, excitations in the 
Moore-Read state in a fractional quantum Hall system obey
non-abelian statistics \cite{MR,NW,Ivanov}.  
Non-abelian statistics we will show here realizes
this type of statistics in relativistic
field theory.
Although the fractional quantum Hall system
violates Parity and Time-reversal symmetries in 1+2 dimensional quantum theory,
our model given below does not violate them in vacuum.

The system we consider is a Majorana fermion in 1+3 dimensions coupled to
an axion field.
The Lagrangian is 
\begin{eqnarray}
{\cal L}=\frac{i}{2}\bar{\psi}_{\rm M}\gamma^{\mu}\partial_{\mu}\psi_{\rm M}
-\frac{1}{2}\bar{\psi}_{\rm M}\left(\Phi_1+i\gamma_5\Phi_2\right)\psi_{\rm M}, 
\end{eqnarray}
where $\psi_{\rm M}$ denotes the Majorana fermion, and
a complex scalar
filed $\Phi=\Phi_1+i\Phi_2$ with a non-zero vacuum expectation value
$|\Phi|=\Phi_0$ denotes the axion filed. 
Here we use a convention $\eta_{\mu\nu}={\rm
diag}(+,-,-,-)$ and define $\gamma^{\mu}$ and $\gamma_5$ as 
\begin{eqnarray}
\gamma^{\mu}=
\left(
\begin{array}{cc}
0&\sigma^{\mu} \\
\bar{\sigma}^{\mu}&0
\end{array}
\right),
\quad
\gamma_5=
\left(
\begin{array}{cc}
1 &0 \\
0&-1
\end{array}\right),
\end{eqnarray}
where 
$\sigma^\mu=\left(1,-\sigma_{i}\right)$,
$\bar{\sigma}^\mu=\left(1,\sigma_{i}\right)$, and $\sigma_i \,(i=1,2,3)$
are the Pauli matrices.
The Majorana condition for $\psi_{\rm M}$ is given by
$i\gamma^2\psi_{\rm M}^{*}=\psi_{\rm M}$.  
This system has an axial U(1) symmetry,
\begin{eqnarray}
\psi_{\rm M}\rightarrow e^{i\gamma_5\theta}\psi_{\rm M},
\quad 
\Phi\rightarrow e^{-2i\theta}\Phi,
\label{eq:axial}
\end{eqnarray}
which is spontaneously broken.
As $\pi_1({\rm U(1)})={\bf Z}$, there exists a
topologically stable string solution called axion string in the broken
phase \cite{CH,HN}.  
The energy per unit length of the axion string is logarithmically divergent,  
but, in the cosmologically context, this divergence is naturally cut off by
the separation between the strings of opposite winding number or by
the size of the string loop. 
In the following, we treat only the straight strings parallel to the
$x^3$-axis and neglect the $x^3$-dependence of the system.

Let us first start with an axion string configuration given by
\begin{eqnarray}
\Phi=\Phi_0 f(\rho)e^{i\phi}, 
\label{eq:phisingle}
\end{eqnarray}
where $\rho$ and $\phi$ are
\begin{eqnarray}
x^1=\rho\cos\phi,
\quad
x^2=\rho\sin\phi.
\end{eqnarray}
The function $f(\rho)$ vanishes on the core of the string, and
approaches to $f(\infty)=1$ far away from the core. 
On the string, there exists one fermionic zero mode \cite{CH}.
The zero mode $u_0$ satisfies the Dirac equation with zero energy,
\begin{eqnarray}
\left(\begin{array}{cc}
-i\sigma_i\partial_i &\Phi^{*} \\
\Phi  &i\sigma_i\partial_i    
\end{array}
\right) u_0=0.
\end{eqnarray}
The solution is given by
\begin{eqnarray}
u_0=C
\left(\begin{array}{c}
0 \\
1+i\\
1-i\\
0      
\end{array}
\right)\exp\left[-\Phi_0\int_0^{\rho}d\sigma f(\sigma)\right],
\label{eq:zeromode}
\end{eqnarray}
where $C$ is a constant. The phase of $C$ can be chosen arbitrary so we
take that $C$ is real so as to satisfy $i\gamma^2u_0^{*}=u_0$. 

On the $n$ strings configuration, there exist $n$ fermionic zero modes
\cite{Semenoff}.      
When the stings are well-separated, the axion filed is approximately given by 
\begin{eqnarray}
\Phi=\Phi_0\prod_{i=1}^{n} f(\rho(x-X_i))e^{i\sum_{i=1}^{n}\phi(x-X_i)}.
\label{eq:phimulti}
\end{eqnarray}
Here $X_i$ denotes the position of the $i$-th string, and 
$\rho(x-X_i)$ and $\phi(x-X_i)$ are given by
\begin{eqnarray}
x^1-X^1_i=\rho(x-X_i)\cos\phi(x-X_i),
\quad
x^2-X^2_i=\rho(x-X_i)\sin\phi(x-X_i).
\end{eqnarray}
We choose $\phi$ so as to satisfy $0\le\phi<2\pi$ in the
following.
Noting that the axion filed (\ref{eq:phimulti}) near the $i$-th string
reduces to (\ref{eq:phisingle}) up to phase factors,
we can write the zero mode on the $i$-th string approximately,
\begin{eqnarray}
u_0^{(i)}(x)=e^{-\frac{i\gamma_5}{2}\sum_{j\neq i}\phi(X_i-X_j)}u_0(x-X_i), 
\label{eq:zeromodemulti}
\end{eqnarray}
where $u_0$ in the right-hand side is the function (\ref{eq:zeromode}). 
This mode $u_0^{(i)}$ also satisfies $i\gamma^2u_0^{(i)*}=u_0^{(i)}$.

Now quantize the system of the $n$ well-separated strings
semi-classically. 
We expand the Majorana filed $\psi_{\rm M}$ by the normalized energy
eigenfunctions $u_{p}$ which satisfy 
\begin{eqnarray}
\left(
\begin{array}{cc}
-i\sigma_i\partial_i &\Phi^{*} \\
\Phi &i\sigma_i\partial_i
\end{array}
\right)
u_{p}
=E_pu_{p},
\end{eqnarray}
and
\begin{eqnarray}
\int dx u_p^{\dagger}(x)u_{p'}(x)=\delta_{p,p'}. 
\end{eqnarray}
If $u_{p}$ is a positive energy eigenfunction, then $i\gamma^2u^{*}_{p}$
becomes a negative energy eigenfunction with energy $-E_p$. So 
the Majorana field $\psi_{\rm M}$ is expanded as
\begin{eqnarray}
\psi_{\rm M}=\sum_{E_p>0}b_p u_{p} e^{-iE_px_0}
+\sum_{E_p>0}b_p^{\dagger}(i\gamma^2u_{p}^*)e^{iE_px_0} 
+\sum_{i=1}^{n}c_0^{(i)}u_0^{(i)}.
\end{eqnarray}
Because of the Majorana condition $i\gamma^2\psi_{\rm M}^*=\psi_{\rm
M}$, the coefficients of the negative modes $i\gamma^2u_p^{*}$ are
hermitian conjugate of that of the positive modes.
Also the same condition implies that the coefficients
of the zero modes satisfy the 1+1 dimensional Majorana condition,
\begin{eqnarray}
\quad c_0^{(i)}=c_0^{(i)\dagger},
\quad
(i=1,\cdots,n).
\label{eq:majoranaweyl}
\end{eqnarray}
Since fermionic zero modes on axion strings are 1+1 dimensional Weyl
fermions \cite{CH}, the low energy effective theory of our model is
written by Majorana-Weyl fermions on the world sheets of the strings.  

From the anti-commutation relations of $\psi_{\rm M}$, we obtain
\begin{eqnarray}
\{b_p,b_{p'}^{\dagger}\}=\delta_{p,p'},
\quad 
\{b_p,b_{p'}\}=0,
\end{eqnarray}
and
\begin{eqnarray}
\{c_0^{(i)},c_0^{(j)}\}=\delta^{i,j}.
\label{eq:c0anticom}
\end{eqnarray}
The commutation relation of  $c_0^{(i)}$ is the Clifford algebra of SO($n$). 

The lowest state of the $n$ axion strings $|X_1,\cdots,X_{n}\rangle$
satisfies
\begin{eqnarray}
b_{p}|X_1,\cdots,X_{n}\rangle
=0,
\label{eq:bcond}
\end{eqnarray} 
since $b_{p}$'s and $b_{p}^{\dagger}$'s are the annihilation and creation
operators of the states with $E_p>0$.
But the action of $c_0^{(i)}$ to the lowest state is not
determined from the consideration of energy: 
The eigenfunctions given by Eq.(\ref{eq:zeromodemulti}) are not exact,
but the index theorem
\cite{Semenoff} implies that the energies are exactly zero.
The only one can make is that the
states provide a representation of the algebra
(\ref{eq:c0anticom}) \cite{JR}. 

The only irreducible representation of the Clifford algebra is the
spinor representation.
When $n=2N$, we introduce the
following operators,  
\begin{eqnarray}
d_0^{(I)}=\frac{1}{\sqrt{2}}\left(c_0^{(2I-1)}+ic_0^{(2I)} \right),
\quad
(I=1,\cdots,N),
\end{eqnarray}
which satisfy
\begin{eqnarray}
\{d_0^{(I)},d_0^{(J)\dagger}\}=\delta^{I,J}, 
\quad
\{d_0^{(I)},d_0^{(J)}\}=\{d_0^{(I)\dagger},d_0^{(J)\dagger}\}=0.
\end{eqnarray}
The spinor representation can be obtained by
\begin{eqnarray}
d_0^{(I)}|X_1,\cdots,X_{n}\rangle =0,
\quad
(I=1,\cdots,N).
\label{eq:d0cond}
\end{eqnarray}
When $n=2N+1$, in addition to Eq.(\ref{eq:d0cond}), we impose that 
\begin{eqnarray}
c^{(2N+1)}_0|X_1,\cdots,X_{2N+1}\rangle
=\frac{1}{\sqrt{2}}|X_1,\cdots,X_{2N+1}\rangle . 
\end{eqnarray}
This construction of the spinor representation is equivalent to  
the standard one in which the Clifford algebra is given by the tensor
products of the Pauli
matrices: 
\begin{eqnarray}
c_0^{(1)}&=&\frac{1}{\sqrt{2}}
\sigma_1\otimes\sigma_3\otimes\cdots\otimes\sigma_3, 
\nonumber\\
c_0^{(2)}&=&\frac{1}{\sqrt{2}}
\sigma_2\otimes\sigma_3\otimes\cdots\otimes\sigma_3, 
\nonumber\\
&\vdots&
\nonumber\\
c_0^{(2I-1)}&=&\frac{1}{\sqrt{2}}1\otimes\cdots\otimes 1\otimes
\sigma_1\otimes\sigma_3\otimes\cdots\otimes\sigma_3,
\nonumber\\
c_0^{(2I)}&=&\frac{1}{\sqrt{2}}1\otimes\cdots\otimes 1\otimes
\sigma_2\otimes\sigma_3\otimes\cdots\otimes\sigma_3,
\nonumber\\
&\vdots&
\nonumber\\
c_0^{(2N-1)}&=&\frac{1}{\sqrt{2}}
1\otimes\cdots\otimes 1\otimes\sigma_1,
\nonumber\\
c_0^{(2N)}&=&\frac{1}{\sqrt{2}}
1\otimes\cdots\otimes 1\otimes\sigma_2,
\nonumber\\
c_0^{(2N+1)}&=&\frac{1}{\sqrt{2}}
\sigma_3\otimes\cdots\otimes\sigma_3, 
\end{eqnarray}
where $c_0^{(2I-1)}$ and $c_0^{(2I)}$ have $I-1$ 1's to the left and
$N-I$ $\sigma_3$'s to the right of $\sigma_1$ and $\sigma_2$, respectively.
The state $|X_1,\cdots,X_n\rangle$ is 
\begin{eqnarray}
|X_1,\cdots,X_n\rangle=|+\rangle\otimes\cdots\otimes|+\rangle
\end{eqnarray} 
with $\sigma_3|\pm\rangle=\pm|\pm\rangle$ in this
tensor product representation.

\begin{figure}[h]
\begin{center}
\includegraphics[width=5cm]{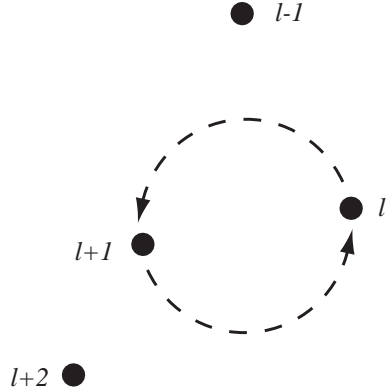}
\end{center} 
\caption{Interchange of the $l$-th string and the $l+1$-th string.}
\label{fig:interchange}
\end{figure}

To examine statistics of this system,  we interchange the $l$-th string and the
$l+1$-th string adiabatically.
Without loss of generality, we assume
that $0\le\phi(X_l-X_{l+1})<\pi$ in the following.  
We also assume that the strings are interchanged counterclockwise with
no other strings between them.  
(See Fig.\ref{fig:interchange}.)   
In this process, the axion filed $\Phi$ starts at Eq.(\ref{eq:phimulti})
and gradually changes then finally returns to the same value
(\ref{eq:phimulti}).
Generally, the wave function $u_{p}$ with $E_p>0$
acquires the (non-abelian) Berry phase $\gamma$ in this process,
\begin{eqnarray}
u_{p}\rightarrow\sum_{E_{p'}=E_p>0} (e^{i\gamma})_{p,p'}u_{p'}. 
\end{eqnarray} 
As $b_p$ is given by
\begin{eqnarray}
b_p=\int dx u_{p}^{\dagger}(x)\psi_{M}(0,x),
\end{eqnarray}
then
\begin{eqnarray}
b_{p}\rightarrow \sum_{E_{p'}=E_p}b_{p'}(e^{-i\gamma})_{p',p}.
\end{eqnarray}
The annihilation operators $b_{p}$'s do not transform to the
creation operators $b_{p}^{\dagger}$'s,
so Eq.(\ref{eq:bcond}) holds after the interchange.

For the zero modes, the transformation law under the interchange can be
calculated explicitly from Eq.(\ref{eq:zeromodemulti}):
The phases of the zero modes $u_0^{(l)}$ and $u_0^{(l+1)}$ transform as
\begin{eqnarray}
\phi(X_l-X_{l+1})\rightarrow\phi(X_{l+1}-X_l),
\quad
\phi(X_{l+1}-X_l)\rightarrow\phi(X_l-X_{l+1})+2\pi,
\end{eqnarray}
so we find 
\begin{eqnarray}
u_0^{(l)}\rightarrow u_0^{(l+1)},
\quad
u_0^{(l+1)}\rightarrow -u_0^{(l)}.
\end{eqnarray}
Using the equations
\begin{eqnarray}
c_0^{(l)}=\int dx u_0^{(l)\dagger}(x)\psi_{M}(0,x), 
\quad
c_0^{(l+1)}=\int dx u_0^{(l+1)\dagger}(x)\psi_{M}(0,x), 
\end{eqnarray}
we obtain
\begin{eqnarray}
c_0^{(l)}\rightarrow c_0^{(l+1)},
\quad
c_0^{(l+1)}\rightarrow -c_0^{(l)}.
\end{eqnarray}
For $i\neq l$ and $l+1$, the phase of the zero
mode $u_0^{(i)}$ does not change by the interchange, 
so $u_0^{(i)}$ transforms trivially
\begin{eqnarray}
u_0^{(i)}\rightarrow u_0^{(i)},
\quad
(i\neq l,l+1),
\end{eqnarray}
and
\begin{eqnarray}
c_0^{(i)}\rightarrow c_0^{(i)},
\quad (i\neq l,l+1). 
\end{eqnarray}

It is remarkable here that these transformation affects
the condition on the state $|X_1,\cdots,X_n\rangle$ given by
Eq.(\ref{eq:d0cond}).  
By an interchange of the $2I$-th and the $2I+1$-th strings, 
we find 
\begin{eqnarray}
d_0^{(I)}&=&\frac{1}{\sqrt{2}}\left(c_0^{(2I-1)}+i c_0^{(2I)}\right)
\nonumber\\
&\rightarrow& \frac{1}{\sqrt{2}}\left(c_0^{(2I-1)}+ic_0^{(2I+1)}\right)=
\frac{1}{2}\left(d_0^{(I)}+d_0^{(I)\dagger}
+id_0^{(I+1)}+id_0^{(I+1)\dagger}\right),
\\
d_0^{(I+1)}&=&\frac{1}{\sqrt{2}}\left(c_0^{(2I+1)}+i c_0^{(2I+2)}\right)
\nonumber\\
&\rightarrow& \frac{1}{\sqrt{2}}\left(-c_0^{(2I)}+ic_0^{(2I+2)}\right)
=\frac{1}{2}\left(id_0^{(I)}-id_0^{(I)\dagger}
+d_0^{(I+1)}-d_0^{(I+1)\dagger}\right).
\end{eqnarray}
Thus the condition (\ref{eq:d0cond}) is not satisfied
after the interchanges. 
The state after the interchange must vanish by the action of the
right hand side of the above equations, 
so we can write down the following transformation law,
\begin{eqnarray}
&&|X_1,\cdots, X_{n}\rangle
\nonumber\\
&&\rightarrow  
\frac{e^{i\theta_{2I,2I+1}}}{\sqrt{2}}\left(
|X_1,\cdots,X_{n}\rangle
-id_0^{(I)\dagger}d_0^{(I+1)\dagger}
|X_1,\cdots,X_{n}\rangle
\right),
\end{eqnarray}
where $\theta_{2I,2I+1}$ is a real constant.

Non-abelian properties of the transformations above can be easily seen by 
considering an interchange of $2I$-th and $2I+1$-th strings
and that of $2I-1$-th and $2I$-th strings simultaneously.
By an interchange of $2I-1$-th and $2I$-th strings, $d_0^{(I)}$
transforms as 
\begin{eqnarray}
d^{(I)}_0&=&\frac{1}{\sqrt{2}}\left(c_0^{(2I-1)}+ic_0^{(2I)}\right)
\nonumber\\
&\rightarrow&\frac{1}{\sqrt{2}}\left(c_0^{(2I)}-ic_0^{(2I-1)}\right) 
=-id_0^{(I)},
\end{eqnarray}
so the lowest state transforms as
\begin{eqnarray}
|X_1,\cdots,X_{n}\rangle
\rightarrow 
e^{i\theta_{2I-1,2I}}|X_1,\cdots,X_{n}\rangle,
\end{eqnarray}
where $\theta_{2I-1,2I}$ is a real constant.
Thus if we interchange $2I-1$-th and $2I$-th strings first then
interchange $2I$-th and $2I+1$-th strings, we obtain
\begin{eqnarray}
&&|X_1,\cdots,X_{n}\rangle
\nonumber\\
&&\rightarrow\frac{e^{i\theta_{2I-1,2I}+i\theta_{2I,2I+1}}}{\sqrt{2}}
\left(|X_1,\cdots,X_{n}\rangle
-id_0^{(I)\dagger}d_0^{(I+1)\dagger}|X_1,\cdots,X_{n}\rangle\right).
\end{eqnarray}
On the other hand, if we interchange $2I$-th and $2I+1$-th strings first
then interchange $2I-1$-th and $2I$-th strings, we obtain
\begin{eqnarray}
&&|X_1,\cdots, X_{n}\rangle
\nonumber\\
&&\rightarrow
\frac{e^{i\theta_{2I-1,2I}+i\theta_{2I,2I+1}}}{\sqrt{2}}\left(
|X_1,\cdots,X_{n}\rangle
+d_0^{(I)\dagger}d_0^{(I+1)\dagger}
|X_1,\cdots,X_{n}\rangle
\right).
\end{eqnarray}
Therefore, the interchange of $2I-1$-th and $2I$-th
strings does not commute with that of $2I$-th and $2I+1$-th strings.
The strings obeys non-abelian statistics.

The interchanges of the $l$-th and $l+1$-th strings can be
summarized by the following unitary operator $T_{l}$: 
\begin{eqnarray}
T_{l}=e^{i\alpha_{l,l+1}}\cdot 
e^{-i\sum_{E_p=E_{p'}}\gamma_{p,p'}b_pb_{p'}^{\dagger}}
\cdot
e^{i\frac{\pi}{4}\Sigma_{l,l+1}}.
\end{eqnarray}
Here $\Sigma_{l,l+1}=i[c_0^{(l)},c_0^{(l+1)}]$ is the generator of SO($n$)
rotation in $l-(l+1)$ plane, and $\alpha_{l,l+1}$ is a real constant.
The operator ${\cal O}(=c_0^{(i)}, b_p)$ transforms as ${\cal O}\rightarrow
T_{l}{\cal O}T_{l}^{\dagger}$, and the state $|\,\,\, \rangle$
transforms as $|\,\,\,\rangle\rightarrow T_{l}|\,\,\,\rangle$. 
The interchange operator $T_{l}$ is the same as that of the spinor
braiding statistics found in a fractional quantum Hall system \cite{NW,Ivanov}
up to a phase factor and the massive mode contributions. 
It is easily shown that $T_{l}$'s are generators of the braid group
\cite{Kauffman}. 
\begin{eqnarray}
&&T_l T_m T_l=T_m T_l T_m,
\quad (|l-m|=1), 
\\   
&&T_l T_m=T_m T_l,
\quad (|l-m|>1).
\end{eqnarray}

In the above, we have considered a model with a spontaneously broken
global U(1) symmetry (\ref{eq:axial}), but 
a gauged version of our model is also possible.
The gauged U(1) symmetry
appears to be anomalous when only the fermion spectrum is considered,
but the anomaly can be cancelled by introducing
an additional anti-symmetric field $B_{\mu\nu}$ and using a variety of
the Green-Schwartz mechanism \cite{HN}.  
This ``anomalous'' U(1) symmetry may arise naturally
in superstring compactification \cite{DSW}. 
Strings in the gauged version of our model
also obey non-abelian statistics by the similar mechanism given above.

It has been shown that unusual statistics is possible for
string loops in 1+3 dimensions \cite{BMSS}.
It is plausible that the axion strings we considered here realize this
statistics non-trivially.

\vspace{5ex}

The author would like to thank M. Shibata and S. Yahikozawa for discussions.
This work was supported in part by the Grant-in-Aid for Scientific
Research No.14740158.


\begin{thebibliography}{99}
\def\J#1#2#3#4{{\sl #1} {\bf #2} (#3) #4}
\def\PL{Phys. Lett.}
\def\NP{Nucl. Phys.}
\def\PRL{Phys. Rev. Lett.}
\def\PR{Phys. Rev.}

\bibitem{JR}R.~Jackiw and C.~Rebbi,
\J{\PR}{D13}{1976}{3398}.
\bibitem{tHooft}G.'t~Hooft,
\J{\PRL}{37}{1976}{8}.
\bibitem{Witten}E.~Witten,
\J{\NP}{B249}{1985}{557}.
\bibitem{Rubakov}V.A.~Rubakov,
\J{\NP}{B203}{1982}{311}.
\bibitem{Callan}C.G.~Callan Jr.,
\J{\PR}{D26}{1982}{2058}.
\bibitem{Sato}M.~Sato,
\J{\PL}{B376}{1996}{41}.
\bibitem{WW}F.~Wilczek and Y.S.~Wu,
\J{\PRL}{65}{1990}{13}.
\bibitem{Bucher}M.~Bucher,
\J{\NP}{B350}{1991}{163}.
\bibitem{BvDdW}F.A.~Bais, P.~van Driel, and M.~de Wild Proptius,
\J{\PL}{B280}{1992}{63}; \J{\NP}{B393}{1993}{547}.
\bibitem{LP}H-K.~Lo and J.~Preskill,
\J{\PR}{D48}{1993}{4821} and references therein.
\bibitem{MR}G.~Moore and N.~Read,
\J{\NP}{B360}{1991}{362}. 
\bibitem{NW}C.~Nayak and F.~Wilczek, 
\J{\NP}{B479}{1996}{529}. 
\bibitem{Ivanov}D.A.~Ivanov,
\J{\PRL}{86}{2001}{268}.
\bibitem{CH}C.G.~Callan Jr. and J.A. Harvey, 
\J{\NP}{B250}{1985}{427}.
\bibitem{HN}J.A.~Harvey and S.G.~Naculich,
\J{\PL}{B217}{1989}{231}.
\bibitem{Semenoff}G.W.~Semenoff,
\J{\PR}{D37}{1988}{2838}.
\bibitem{Kauffman}See, e.g., L.H.~Kauffman, {\it Knots and Physics},
	(World Scientific, Singapore, 1991)
\bibitem{DSW}M.~Dine, N.~Seiberg, and E.~Witten
\J{\NP}{B289}{1987}{589}.
\bibitem{BMSS}A.P.~Balachandran, G.~Marmo, B.S.~Skagerstam, and A.~Stern,   
{\it Classical Topology and Quantum States}, (World Scientific,
	Singapore, 1991), chapter 22. 
\end{thebibliography}
\end{document}